\documentclass[twocolumn, nofootinbib, aps, prd, floats, floatfix, amsmath, amssymb, longbibliography, superscriptaddress, preprintnumbers]{revtex4-1} %
\pdfoutput=1
\usepackage[final]{graphicx}
\usepackage{hyperref}
\usepackage{amsmath}
\usepackage{bbm}
\usepackage{bm}
\usepackage{amsfonts}
\usepackage{amssymb}
\usepackage{latexsym}
\usepackage{graphicx}
\usepackage[english]{babel}
\usepackage{multirow}
\usepackage{float}
\usepackage{url}
\usepackage{slashed}
\usepackage{xcolor} 
\usepackage[utf8]{inputenc}
\usepackage{stmaryrd} 
\usepackage{enumitem}
\usepackage{hyperref}
\usepackage{cleveref}
\usepackage{siunitx}
\usepackage{verbatim}

\usepackage{amsthm}
\usepackage{ulem}

\definecolor{colorML}{rgb}{.2,.7,.7}
\definecolor{colorZL}{rgb}{.2,.2,.7}
\definecolor{colorWC}{rgb}{.7,.2,.7}

\begin{document}

\title{Dark matter transient annihilations in the early Universe}

\author{Katsuya Hashino}
\email{hashino@pku.edu.cn}
\affiliation{Center for High Energy Physics, Peking University, Beijing 100871, China}

\author{Jia Liu}
\email{Corresponding author: jialiu@pku.edu.cn}
\affiliation{School of Physics and State Key Laboratory of Nuclear Physics and Technology, Peking University, Beijing 100871, China}
\affiliation{Center for High Energy Physics, Peking University, Beijing 100871, China}

\author{Xiao-Ping Wang}
\email{Corresponding author: hcwangxiaoping@buaa.edu.cn}
\affiliation{School of Physics, Beihang University, Beijing 100083, China}
\affiliation{Beijing Key Laboratory of Advanced Nuclear Materials and Physics, Beihang University, Beijing 100191, China}

\author{Ke-Pan Xie}
\email{Corresponding author: kepan.xie@unl.edu}
\affiliation{Department of Physics and Astronomy, University of Nebraska, Lincoln, NE 68588, USA}

\begin{abstract}
The cosmological evolution can modify the dark matter (DM) properties in the early Universe to be vastly different from the properties today. 
Therefore, the relation between the relic abundance and the DM constraints today needs to be revisited.
We propose novel \textit{transient} annihilations of DM which helps to alleviate the pressure from DM null detection results.
As a concrete example, we consider the vector portal DM and focus on the mass evolution of the dark photon. 
When the Universe cools down, the gauge boson mass can increase monotonically and go across several important thresholds; opening new transient annihilation channels in the early Universe.
Those channels are either forbidden or weakened at the late Universe which helps to evade the indirect searches.
In particular, the transient resonant channel can survive direct detection (DD) without tuning the DM to be half of the dark photon mass and can be soon tested by future DD or collider experiments.
A feature of the scenario is the existence of a light dark scalar.
\end{abstract}
\maketitle

\section{Introduction}
The Weakly Interacting Massive Particle (WIMP) paradigm provides an elegant explanation to dark matter (DM) via the freeze-out mechanism. It suggests that the DM particle has weak scale couplings to the Standard Model (SM) particles. This implies sizable scattering rates between local DM and nucleons~\cite{LUX:2016ggv, CDEX:2019hzn, XENON:2018voc, PandaX:2021osp}, residual DM annihilation today in galaxies~\cite{AMS:2014xys, AMS:2014bun, Fermi-LAT:2015att, Fermi-LAT:2016uux, DAMPE:2017fbg}, and DM production at colliders~\cite{CMS:2018yfx, ATLAS:2019cid,  ATLAS:2021kxv, CMS:2021far}. However, the null results from the above experiments cast doubt on the WIMP paradigm.

One of the most common benchmark WIMP model is the vector portal DM; in which the DM fermion $\psi$ interacts with the SM particles through kinetic mixing of the $U(1)_d$ dark photon $A'$
and SM photon~\cite{Holdom:1985ag}. The ratio $r_0\equiv m_{A'} /m_\psi $ classifies the parameter space into different regions. For $r_0 <1$, the classic secluded annihilation to dark photon pair is kinematically allowed~\cite{Pospelov:2007mp}. For $1 \lesssim r_0\lesssim 2$, there are new channels allowing secluded annihilations~\cite{DAgnolo:2015ujb, Cline:2017tka, Fitzpatrick:2020vba, Xing:2021pkb, Fitzpatrick:2021cij}. For $r_0 >2$, the DM pair will annihilate into SM particles through $s$-channel $A'$ mediation; thus, there is a direct connection between the relic abundance and nucleon scattering cross-section. For DM mass $> 10$ GeV, most of the  parameter space is already ruled out by direct detection (DD). The exception being the cases of $r_0 \approx 2$ (the fine-tuned $s$-channel resonant region) or inelastic DM (with small mass splitting)~\cite{Cirelli:2016rnw, Liu:2017lpo, Tucker-Smith:2001myb, Alekhin:2015byh, Battaglieri:2017aum, BDX:2017jub, Berlin:2018bsc, LDMX:2018cma, Tsai:2019buq}. Light DM can avoid DD, but is still subject to constraints from cosmic microwave background (CMB) measurements~\cite{Slatyer:2015kla, Cirelli:2016rnw} and the intensity frontier experiments 
\cite{Alekhin:2015byh, Battaglieri:2017aum}. Therefore, the vector portal DM model is severely constrained by the experiments.

In this paper, we point out that the DM evolution can be deeply affected by the thermal history of the Universe; hence, the above constraints cannot be trivially applied. More specifically, the $U(1)_d$ is restored 
when the cosmic temperature is very high. If the $U(1)_d$ breaking is through a second-order phase transition, then, as the Universe cools down, the $A'$ mass will scan from zero to today's zero temperature value. For $r_0>2$, such a ``mass scanning'' will open transient secluded annihilations and $s$-channel resonant annihilation which help evade current DD, indirect detection, and collider searches. This scenario is testable by near future DM experiments and can serve as a viable variant of the WIMP model. 

There have been prior studies on the effects cosmological evolution have on DM or scalar mediator mass, stability, interaction couplings, and annihilation channels~\cite{Cohen:2008nb, Baker:2016xzo, Kobakhidze:2017ini, Baker:2017zwx, Baker:2018vos, Hektor:2018esx, Bian:2018mkl, Bian:2018bxr, Kobakhidze:2019tts, Heurtier:2019beu, Darme:2019wpd, Davoudiasl:2019xeb, DeRomeri:2020wng, Jaramillo:2020dde, Croon:2020ntf, Nakayama:2021avl, Batell:2021ofv}. In our scenario, we for the first time focus on the vector mediator whose mass is significantly affected during the freeze-out. The DM mass and its couplings are not affected at the freeze-out temperature. Moreover, the Higgs boson responsible for the $U(1)_d$ breaking has to be much lighter than $A'$; a feature of this scenario.
\\

\section{Model}
The vector portal DM model has the following Lagrangian
\begin{align}
\mathcal{L}_d=\bar{\psi} \left(i\slashed{D} - m_\psi \right) \psi - \frac{1}{4}F'_{\mu\nu}F'^{\mu\nu} + \epsilon e A^{'}_\mu J^\mu_{\rm em},
\end{align}
where $D_\mu\equiv\partial_\mu-ig_dA'$, and $J^\mu_{\rm em} $ is the SM electromagnetic current. The coupling between $J^\mu_{\rm em}$ and $A'$ comes from the kinetic mixing with the photon field strength. The $U(1)_d$ is spontaneously broken by a complex scalar $\Phi=(\phi+i\eta)/\sqrt{2}$ with the potential
\begin{align}
	V(\Phi)=\mu_d^2 |\Phi|^2+\lambda_d |\Phi|^4.
	\label{eq:Lag_phi}
\end{align}
Provided that $\mu_d^2<0$, the scalar field obtains a vacuum expectation value (VEV) $\langle \phi\rangle=v_d\equiv \sqrt{-\mu_d^2/\lambda_d}$, and hence, $A'$ acquires a mass $m_{A'}=g_dv_d$. The real part of the scalar $\phi$ also obtains a mass $m_\phi=\sqrt{2\lambda_d}v_d$.

The scalar potential in Eq.~(\ref{eq:Lag_phi}) receives corrections in the early Universe. For example, if there is a light scalar $S$ coupling to $\Phi$ via $\lambda_{\phi S}|\Phi|^2S^2$, a thermal mass term $\lambda_{\phi S}T^2\phi^2/12$ can be induced. Another possible correction comes from the gravitational coupling $\xi_\phi R|\Phi|^2/2$, where $R$ is the Ricci curvature scalar. As $R/H^2\sim10^{-2}$--$10^{-1}$ at temperature between 100 MeV and 10 GeV (with $H$ being the Hubble constant)~\cite{Croon:2020ntf,Davoudiasl:2004gf,Caldwell:2013mox}, the portal coupling provides a $T^4$ correction to the scalar mass term.
Moreover, corrections like $m_*H\Phi^2$ or $H^2\Phi^2$ can also arise from the flat directions in supersymmetric models~\cite{Dine:1995uk, Dine:1995kz}.
In general, the potential in the early Universe can be written as
\begin{align}
	\mu_d^2 (T) = \mu_{d,0}^2 + c_\phi T^n,
	\label{eq:mudparameter}
\end{align}
where the dimension of the coefficient $c_\phi$ is $[{\rm Energy}^{(2-n)}]$ with $n=2$ or 4. The scalar VEV varies with the temperature as,
\begin{align}
v_d^2(T)=\begin{cases}~0 \quad &T>T_\phi  \\ 
	~v_{d,0}^2-c_\phi T^n/\lambda_d \quad &T<T_\phi\end{cases} ,
\end{align}
where $T_\phi=(\lambda_dv_{d,0}^2/c_\phi)^{1/n}$ is a temperature at which the second-order phase transition for $U(1)_d$ breaking starts. The evolution of $m_{A'}$ can be derived immediately, i.e.,
\begin{align}
m_{A'}^2(T) = \begin{cases} ~0\quad &T>T_\phi,  \\
	~m_{A',0}^2 - \kappa m_\psi^2\left(\frac{T}{m_\psi}\right)^n \quad & T<T_\phi
	\end{cases} ,
\label{eq:Apmass}
\end{align}
where $m_{A',0} = g_d v_{d,0}$ is the mass at zero temperature and $\kappa=m_\psi^{n-2}c_\phi g_d^2/\lambda_d$ is a model dependent dimensionless constant. Later, we will see that $\kappa$ is required to be large.

The mass of $\phi$ is also temperature-dependent,
\begin{align}
	\label{eq:phimass}
	m_\phi^2(T) = \begin{cases} ~ \mu_{d,0}^2+c_\phi T^n  \quad  & T>T_\phi  	 \\
		~m_{A'}^2(T)\times\left( 2 m_\psi^{n-2}c_\phi/\kappa\right) \quad & T<T_\phi
	\end{cases}.
\end{align}
Since $\kappa$ is large, a scalar $\phi$ much lighter than $A'$ is a feature of our model. More specifically, for $n=2$, we are interested in $\kappa\sim10^4$ with $c_\phi\sim1/12$, and hence, the mass of $\phi$ is sub-GeV. To avoid constraints from cosmological observations, a small Higgs portal coupling $\lambda_{h\phi}|H|^2|\Phi|^2$ is assumed to allow $\phi$ to decay to a pair of SM light fermions before Big Bang Nucleosynthesis. 
For $n=4$, since $c_\phi\sim1/m_{\rm pl}^2$, with $m_{\rm pl}=1.22\times10^{19}$ GeV being the Planck scale, a very small $\lambda_d\sim10^{-38}$ is required for $T_\phi\sim 1$ GeV. Hence, an extremely light $\phi$ with $m_\phi\sim\mathcal{O}(10^{-10}~{\rm eV})$ exists. 
Due to its tiny mass, its decay to diphoton via SM Higgs mixing is too slow comparing with the Hubble time scale, therefore making it a stable particle. Such ultralight $\phi$ can exist as dark radiation, leaving impacts on the cosmic large scale structure of the Universe~\cite{Tang:2016mot}.
The smallness of $\lambda_d$, or in other words, the flatness of the potential Eq.~(\ref{eq:Lag_phi}) can be achieved by embedding the model into either a spontaneously broken global symmetry with $\Phi$ as the pseudo-Nambu-Goldstone boson or a supersymmetric model with $\Phi$ as the moduli field. 
Although Eq.~(\ref{eq:mudparameter}) seems to be a simplified model, it can be treated as the prototype of a general continuous phase transition in the sense of Taylor expansion around the  critical temperatures in Eq.~(\ref{eq:characteristic}). Therefore, the methodology can apply to the general case for a more complicated potential as described in Appendix~\ref{app:pt}. In addition, the reason $n=2$ case can be a good approximation for the one-loop finite temperature potential including the Coleman-Weinberg potential and the thermal corrections, is given in Appendix~\ref{app:thermal}.

In summary, the cosmological evolution effects on $A'$ is fully encoded in the constant $\kappa$. This serves as an extra free parameter compared to the zero temperature model. Therefore, there are five input parameters in total,
\begin{align}
\left\{m_{\psi}, ~m_{A',0}, ~g_d, ~\epsilon , ~\kappa\right\}.
\end{align}
For convenience, we drop the $(T)$ for the explicit temperature dependence thereafter, e.g., $m_{A'}$ always implies $m_{A'}(T)$. The subscript ``$0$'' denotes the zero temperature values.
\\

\section{Transient annihilations}

For $r_0 = m_{A',0}/m_\psi>2$, as the temperature drops, $m_{A'}$ will inevitably go across several thresholds:
\begin{align}\label{eq:characteristic}
\text{\bf Transient secluded:}~{ (A'A')} & \quad  m_{A'} = m_\psi, \nonumber \\ 
{ (A'\phi)}~ & \quad m_{A'} = 2m_\psi - m_\phi,\nonumber \\
\text{\bf Transient resonant:} \quad { (\bar{f}f)} &\quad m_{A'} = 2m_\psi.
\end{align}
Crossing the first two thresholds open up new transient annihilation channels $\bar{\psi}\psi \to A'A'$ and $\bar{\psi}\psi \to A' \phi$. These are secluded annihilation and thus, can evade the DD limits and the collider constraints. 
The last crossover enables a transient $s$-channel resonance, $\bar{\psi}\psi \to A' \to \bar{f}f$, which greatly enhances the annihilation cross-section. Defining $x \equiv m_\psi /T$, we denote those temperatures as $x_{A'A'}$, $x_{A'\phi}$ and $x_{\rm res}$ respectively.
If they happen to be around the freeze-out temperature $x_{\rm fo} \sim 23$, the relic abundance calculation has to incorporate those transient annihilations. It generally requires a large $\kappa$
\begin{align}
\kappa\gtrsim x_{\rm fo}^n (r_0^2-a),
\label{eq:kappa}
\end{align} 
with $a=1$ (4) for $A'A'$ $(\bar{f}f) $ final states respectively.

\begin{figure*}[htpb]
		\includegraphics[width=0.48\textwidth]{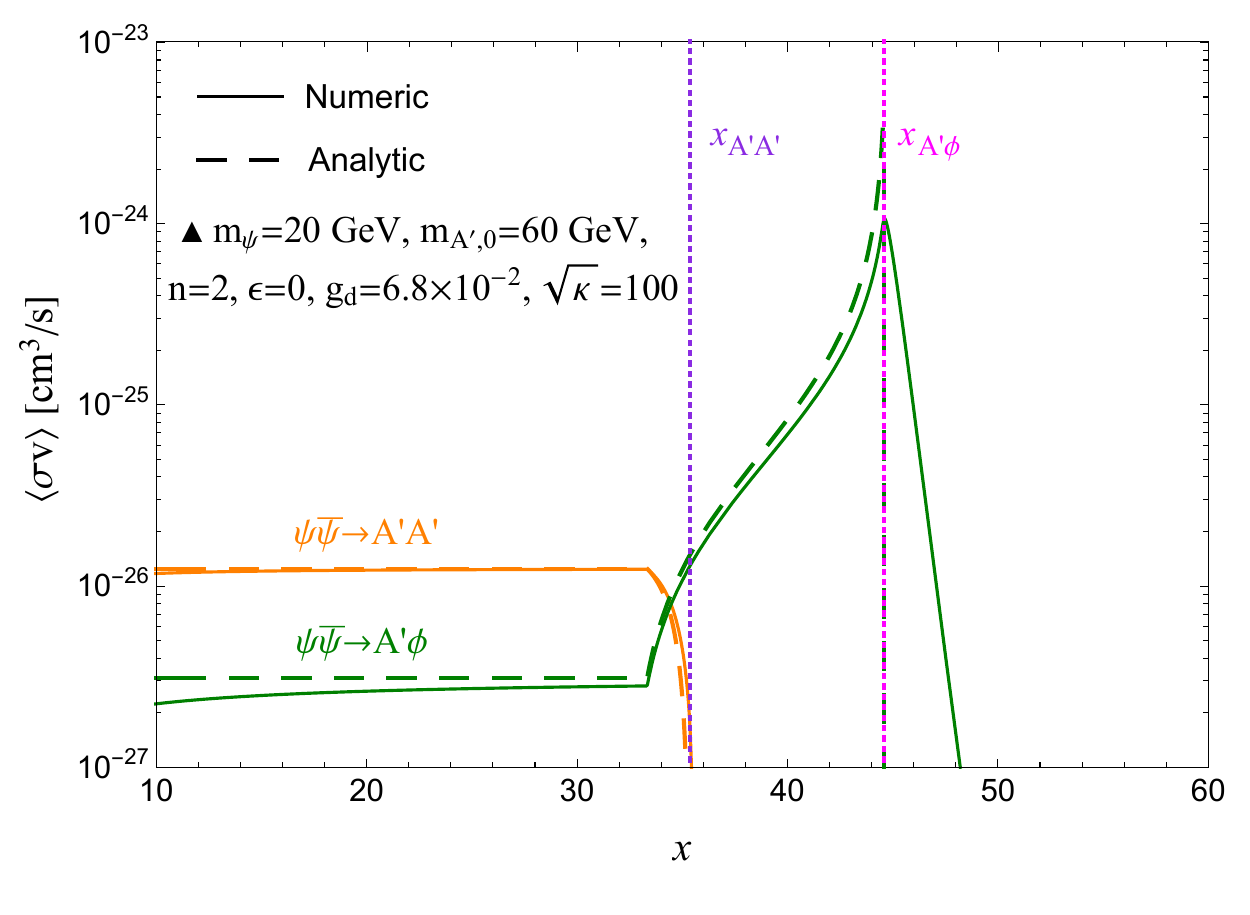} 
		\includegraphics[width=0.48\textwidth]{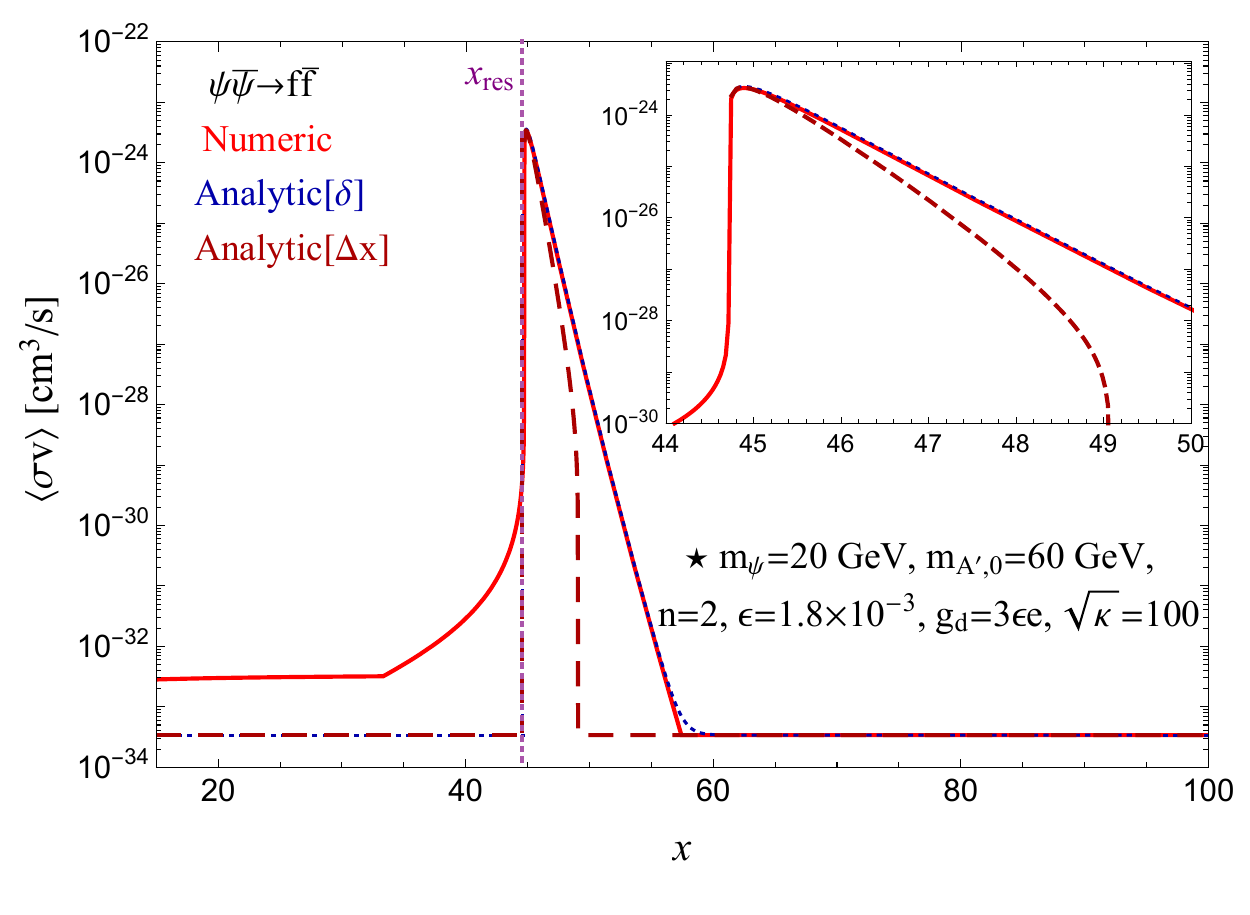}
	\caption{The $\langle\sigma v\rangle$ from analytic and numeric calculations for $A'A'$ and $A'\phi$ in the left panel and for resonant $\bar{f}f$ in the right panel, with $n=2$. The benchmark $\blacktriangle$ has a large $g_d$ and works for transient secluded annihilation, while $\bigstar$ has a small $g_d$ and is for transient secluded annihilation. An inset figure is provided for the right panel near the threshold $x_{\rm res}$.}
	\label{fig:sigmav}
\end{figure*} 

The relevant annihilation cross-sections are
\begin{align}
&	\left\langle \sigma v_{A'A'}\right \rangle \approx ~\frac{g_d^4}{16\pi m_\psi^2} 
	 \left(1 - r^2 \right)^{3/2} 	\left(1 - r^2/2 \right)^{-2} , 
	 \label{eq:AnnApAp}\\
&	\left \langle \sigma v_{A'\phi}\right \rangle
 \approx \sigma v_{A'\phi}=	\frac{g_d^4 \left(-2 \left(q^2-20\right) r^2+\left(q^2-4\right)^2+r^4\right) }{256 \pi  m_\psi^2 \left(r^2-4\right)^2} \nonumber\\
& \times \sqrt{q^4-2 q^2 \left(r^2+4\right)+\left(r^2-4\right)^2}, \nonumber \\
& \approx \frac{g_d^4}{256 \pi m_\psi^2} 
\frac{r^4 + 40 r^2 +16}{\left| 4-r^2 \right|} +\mathcal{O}\left(q^2\right),
	 \label{eq:AnnApPhi}\\
&	\langle\sigma v_{\bar{f}f}\rangle = ~\frac{g_d^2 \epsilon^2 e^2}{6 \pi} \frac{s+ 2 m_\psi^2}{(s-m_{A'}^2)^2 + m_{A'}^2 \Gamma_{A'}^2},
\label{eq:AnnXecRES}\end{align}
where $r \equiv m_{A'}/m_\psi$ and $q \equiv m_\phi/m_\psi$ is a small parameter, $s$ is the total energy square in the center of mass frame and $\Gamma_{A'}$ is the decay width for $A'$ which is also temperature dependent.

For the \textit{transient secluded} channels ($A'A'$ and $A'\phi$), we approximate their thermally averaged values as the $s$-wave part of their cross-section. The cosmological evolution effects on $m_{A'}$ is not affected by this approximation as the thermal average is taken over the DM velocity distribution. 

For $\left \langle \sigma v_{A'\phi}\right \rangle$, the above equation is obtained by expanding over small $m_\phi$ and is used in the analytic relic abundance calculation. 
Since the cross-section is $s$-wave dominant, we use the approximation  $\left \langle \sigma v_{A'\phi}\right \rangle  \approx  \sigma v_{A'\phi}$, where the latter is the cross-section without thermally averaging. In the analytic calculation for DM yield, we have used the simpler form in the second line. While in the numeric calculation, we have used the expression in the first line.
This channel has an accidental resonant enhancement due to small $m_\phi$, but never hits the resonant peak. This occurs near $x_{A' \phi}$, when the factor $|4- r^2|^{-1}$ becomes $m_\psi/(4 m_\phi)$ at the leading order.
Due to this enhancement, the annihilation channel $A' \phi$ dominates over $A'A'$ most of the time.

For the \textit{transient resonant} channel ($\bar{f}f$), one generally performs the thermal average numerically. 
However, to understand the transient resonant annihilation better, we  
simplify its expression for the analytic relic abundance calculation. For narrow width resonances, the resonance peak of Eq.~(\ref{eq:AnnXecRES}) can be approximated by a $\delta$-function. This leads to the following expression for the thermally averaged cross-section,
\begin{align}
	\left\langle \sigma v\right\rangle_{\bar{f}f}^{\rm res}  \approx \frac{g_d^2 \epsilon^2 e^2 (2+r^2)x}{48 \sqrt{2\pi} m_\psi \Gamma_{A'} } \sqrt{r (r^2-4)x}e^{-(r-2)x}.
	\label{eq:annAVE_res}
\end{align}
This is valid for $x > x_{\rm res} = \left( \kappa/(r_0^2 - 4) \right)^{1/n} $ with resonance at $m_{A'}|_{x_{\rm res}}=  2m_{\psi}$. The width of $A'$ can be approximated as 
\begin{align}
	\Gamma_{A'} \approx \frac{m_{A'}}{12 \pi} \left[\epsilon^2 e^2 n_{\rm dof} + 
	g_d^2 \sqrt{1- \frac{4}{r^2}} \left(1+ \frac{2}{r^2}\right)  \right]
	, \label{eq:ApWidth}
\end{align} 
where $n_{\rm dof }\equiv \sum_f Q_f^2 N_f^c = 20/3$ with charge $Q_f$ and color factor $N_f^c$, summed over SM fermions lighter than the top quark. We neglect the SM fermion masses to further simplify the decay width. 
In the small $g_d $ regime, $g_d \lesssim \sqrt{n_{\rm dof}} \epsilon e$, the decay width is dominated by the SM contribution and the cross-section $\left\langle \sigma v\right\rangle_{\bar{f}f}^{\rm res} $ will be proportional to $g_d^2$. While if $g_d \gtrsim \sqrt{n_{\rm dof}} \epsilon e$, the width is controlled by the invisible decay to $\psi\bar\psi$ and the cross-section $\left\langle \sigma v\right\rangle_{\bar{f}f}^{\rm res} $ will be proportional to $\epsilon^2 e^2$. Therefore, small $g_d$ is more interesting for transient resonant annihilation. 

The cross-section $	\left\langle \sigma v\right\rangle_{\bar{f}f}^{\rm res}$ has an exponential penalty factor from Boltzmann suppression but is unsuppressed near the resonance. 
The cross-section can be greatly simplified by taking $x = x_{\rm res} + \Delta x$ and expanding to leading order in $\Delta x \ll 1$. An important step is keeping the exponential term; otherwise, the Boltzmann suppression vanishes. In this limit, we have
\begin{align}
	& \sum_f	\left\langle \sigma v\right\rangle_{\bar{f}f}^{\rm res}    \approx
\frac{3 \sqrt{ n \pi  \Delta x} g_d^2  }{4 m_\psi^2 } e^{-\frac{ n \Delta x}{4}
	\left(r_0^2 -4 \right)} 
\left(r_0^2 -4 \right)^{\frac{n-2}{2 n}} \nonumber \\
& \times	\kappa^{1/n}
\left(1 - \frac{3 g_d^2 \sqrt{n \Delta x}}{4 n_{\rm dof } \epsilon^2 e^2}  \left(r_0^2 -4 \right)^{\frac{n+1}{2n}}\kappa^{\frac{-1}{2n}} \right),
    \label{eq:ff-AVE-simplified}
\end{align}
where the expansion of small $\Delta x$ also implicitly requires small $g_d$, i.e., $g_d \lesssim \sqrt{n_{\rm dof}} \epsilon e$, because in the last term, if $g_d^2 \sqrt{n \Delta x} \gtrsim n_{\rm dof} \epsilon^2 e^2  $ it will be invalid and return a negative result. At leading order, we see the cross-section increases with $\kappa^{-1/n} g_d^2$.

We compare our analytic calculations with the numeric integration in Fig.~\ref{fig:sigmav}.
In the left panel of Fig.~\ref{fig:sigmav}, we show the analytic results agree with the numeric calculation for
both transient secluded channels. For $A'A'$, at small $x$ the $A'$ is massless thus the cross-section is flat.
Near the threshold $x_{A'A'}$ (i.e. $m_{A' } = m_\psi$) both results drops but for different reasons. The analytic result decreases due to phase space, while numeric results also incorporate Boltzmann suppression beyond the threshold.
For $A'\phi$, the analytic result becomes zero  at threshold $x_{A'\phi}$, while the numeric result has a Boltzmann tail from thermally averaging. 
For $x < x_{A'A'}$, the $A'A'$ channel contribution dominates, while for $x > x_{A'A'}$, the accidental resonant $A'\phi$ channel takes over. Both channels together provide the right relic abundance for the benchmark $\blacktriangle$.

In the right panel of Fig.~\ref{fig:sigmav}, we check the calculation for the $\bar{f}f$ resonant channel. We plot the analytic result in Eq.~(\ref{eq:ff-AVE-simplified}) and label it as ``Analytic[$\Delta x$]" (dashed maroon) in the right panel of Fig.~\ref{fig:sigmav}, which is a simplified form after $\Delta x$ expansion. 
The analytic result without the expansion in Eq.~(\ref{eq:annAVE_res}) is labeled as ``Analytic[$\delta$]" (dotted blue line). It is clear that the $\delta$-function approximation for the resonance peak is quite successful comparing with numeric result in red solid line. The simple analytic expression in Eq.~(\ref{eq:ff-AVE-simplified}) deviates from the other two at large $\Delta x$, but such simplification is necessary for the analytic relic abundance calculation.
Near the resonance, the three results agree quite well with each other. Because the integration of $\left\langle \sigma v\right\rangle$ over $x$ returns similar results, the simple expression can lead to a good match for relic abundance with the other two calculations.

\section{Relic abundance}

The DM relic abundance can be obtained by solving the Boltzmann equation. Using the DM yield $Y \equiv n_\psi/s$, with $s$ being the entropy density, one can reformulate the equation as
\begin{align}
		\frac{dY}{dx} = - \sqrt{\frac{\pi g_*}{45}} \frac{m_{\rm pl} m_\psi}{x^2} \left\langle \sigma v\right\rangle \left(Y^2 - Y_{\rm eq}^2\right), 
	\label{eq:Yrelic1}
\end{align}
where $g_{*}$ is the number of effective degree of freedom and $Y_{\rm eq}$ is the yield at equilibrium. The thermally averaged cross-section $\left\langle \sigma v\right\rangle$
includes all DM annihilation channels. The DM relic abundance can be computed numerically;
noting that $m_{A'}$ and $m_\phi$ will change with $x$.

A more strict treatment of the freeze-out in the narrow resonance case and forbidden annihilation can be found in Ref.~\cite{Binder:2021bmg}, where a technique is developed for solving the full Boltzmann equations when the DM particles are not in kinetic equilibrium with the SM particles. The full treatment will not change our qualitative picture here, but quantitatively yield an $\mathcal{O}(1)$ correction to the relic abundance for two reasons. First, in the transient resonant case, we are interested in DM mass much heavier than the SM fermions ($r\equiv m_f/m_\psi\ll1$, except the top quark), thus the influence from the full Boltzmann equation approach is mild (see Fig.~2 of Ref.~\cite{Binder:2021bmg}). 
Second, for $4m_\chi^2 < m_{A'}^2$ ($4m_\chi^2 > m_{A'}^2$) regions, the relic abundance in full approach is larger (smaller) comparing to the standard approach. Since in our scenario, the mass $m_{A'}$ is changing with temperature and will go through both two regions.
Therefore, after integrating over the $m_{A'}$ scanning range, the corrections from two different regions tend to cancel each other. As a result, the full approach should only provide a small correction to our result.

Furthermore, the relic abundance can be computed analytically as 
\begin{align}
		Y^{-1}(x = \infty) 	\approx \int_{x_{\rm fo}}^{\infty} dx \sqrt{\frac{\pi g_*}{45}} \frac{m_{\rm pl}m_\psi}{x^2} \left\langle \sigma v\right\rangle,
		\label{eq:Ym1fin}
\end{align} 
where we have used the approximations that when freeze-out starts, $Y \gg Y_{\rm eq}$ and $Y(x_{\rm fo}) \gg Y(x=\infty)\equiv Y_0$. 

Therefore, the annihilation contribution from $A'A'$ can be obtained by plugging in Eq.~(\ref{eq:AnnApAp}),
\begin{align}
	Y^{-1}_{A'A'} = \frac{g_d^4 g_{*}^{1/2} m_{\rm pl} }{48 \sqrt{5 \pi } m_\psi} F_{A'A'}\left( x, r_0,\kappa \right)\Big|^{x_{\rm A'A'}}_{x_{\rm fo}}.
\end{align} 
Similarly, we obtain the yield inverse $Y^{-1}_{A' \phi}$ using the simplified thermally averaged cross-section in Eq.~(\ref{eq:AnnApPhi}),
\begin{align}
	Y^{-1}_{A'\phi}  = \sqrt{\frac{g_*}{5\pi}}\frac{g_d^4 m_{\rm pl}}{768m_\psi} F_{A'\phi}(x, r_0,\kappa) \Big|^{x_{\rm A'\phi}}_{x_{\rm fo}}.
\end{align} 
The indefinite integration functions $F_{A'A'}$ and $F_{A'\phi}$ are
\begin{widetext}
\begin{align}
	F_{A'A'}(x, r_0,\kappa)  = &\left\{ \sqrt{r_{0}^{2}-2}\left(2 \left(r_{0}^{2}-2\right)  \operatorname {arctanh} \left[\sqrt{1-\frac{\left(r_0^{2}-1\right) x^{2}}{\kappa}}\right] + \frac{\sqrt{\kappa\left(\kappa-\left(r_{0}^{2}-1\right) x^{2}\right)}}{\kappa - \left(r_{0}^{2}-2 \right) x^{2}}\right) \right.\nonumber \\
	& \left. - \left(3 r_{0}^{2}-5\right) \operatorname{arctan}\left[\sqrt{\frac{\left(r_{0}^{2}-2\right)\left(\kappa-\left(r_{0}^{2}-1\right) x^{2}\right)}{\kappa}}\right]  \right\}\times \frac{-2}{\sqrt{\kappa} \left(r_{0}^{2}-2\right)^{3 / 2}} , \\
	F_{A'\phi}(x, r_0,\kappa)  = &
	\frac{ 192}{\sqrt{(r_0^2-4)\kappa} } \operatorname{arctanh}\left(\frac{\sqrt{r_0^2 - 4}}{\sqrt{\kappa}}x \right)+ \frac{r_0^2+44}{x} - \frac{\kappa}{3 x^3},
\end{align}
\end{widetext}
for $n=2$, where we have assumed $g_{*}^{1/2}$ to be a constant. At the threshold $x_{A'A'}$ (i.e. $r=1$),
the function $F_{A'A'}$ becomes zero because the corresponding annihilation cross-section is proportional to high power in $1- r$.
For the $A'\phi$ channel, the function $F_{A'\phi}$ at threshold is non-zero.

For the resonant channel $\bar{f}f$, using the resonant cross-section Eq.~(\ref{eq:ff-AVE-simplified}), one can obtain 
\begin{align}
	Y^{-1}_{\rm res} & \approx \sqrt{\frac{\pi^3g_*}{5}}\frac{g_d^2m_{\rm pl}}{n m_\psi}  
\left(r_0^2 -4 \right)^{\frac{1-n}{n}} \kappa^{-1/n} \nonumber \\
&	\times \left(1 - \frac{3 g_d^2  (r_0^2 - 4)^{1/(2n)} }{\sqrt{\pi } n_{\rm dof} \epsilon^2 e^2 }\kappa^{-1/(2 n)}\right).
	 \label{eq:Yres-short}
\end{align}
At leading order, the DM relic abundance is proportional to
$g_d^{-2}\kappa^{1/n}$, which mildly depends on parameter $\kappa$. 
Together with the lower bound on $\kappa$ in Eq.~(\ref{eq:kappa}), the model does not need fine-tuning comparing with the normal resonant model. For different $n$, one can choose a $\kappa$ to get the same relic abundance. For example, when switching from $n=2$ to $n=4$, one can rescale $\kappa\to\kappa^2/(16(r_0^2 - 4))$ to obtain similar $Y_{\rm res}$.

\begin{figure}
	\centering
	\includegraphics[width=0.98 \columnwidth]{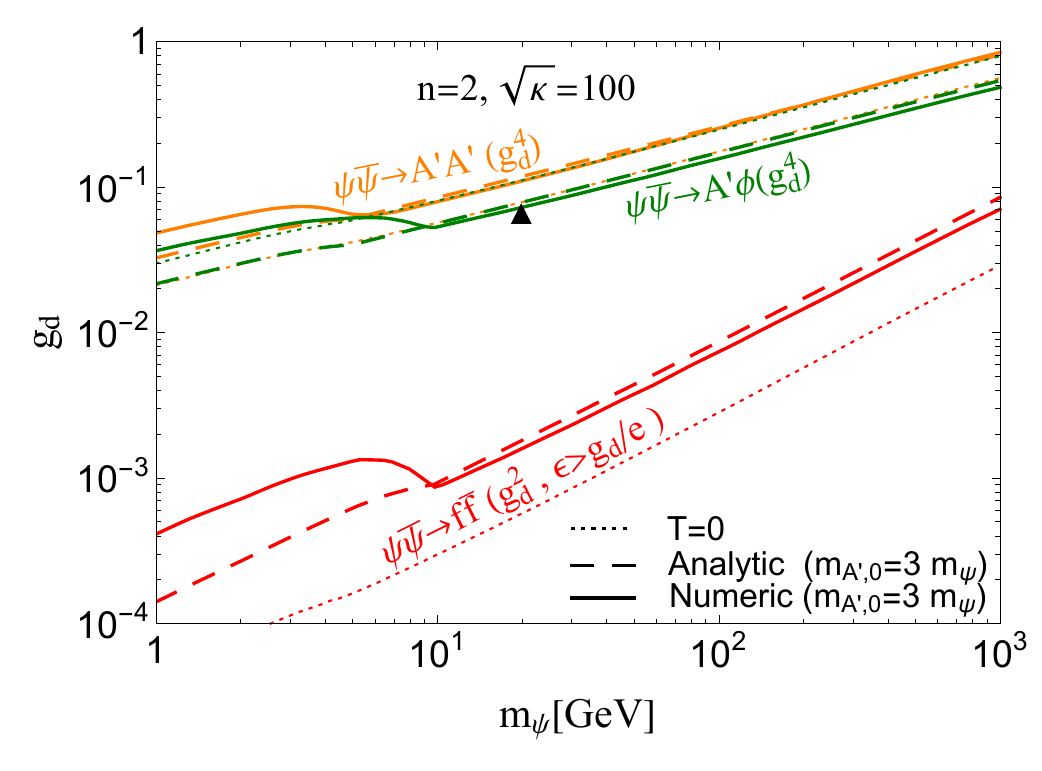} 
	\caption{$g_d$ as a function of $m_\psi$ for transient annihilation channels $A'A'$, $A' \phi$ and $\bar{f}f$, which provides the right relic abundance for $n=2$. The dashed and solid lines are analytic and numeric calculations respectively. The discrepancies between analytic and numeric results for $m_\psi\lesssim10$ GeV  comes from the temperature dependence of $g_*$, which is ignored in the analytic calculation. We also compare them with a set of zero temperature examples (dotted lines) described in the text. The benchmark $\blacktriangle$ has included the contribution from both $A'A'$ and $A'\phi$, and will be described in Fig.~\ref{fig:Y-x-plot}.}
	\label{fig:mDM-gd}
\end{figure} 

Now we show the numerical results. In Fig.~\ref{fig:mDM-gd}, we show the required $g_d$ to obtain the correct relic abundance for each individual annihilation channel $A'A'$, $A'\phi$ and $\bar{f}f$ with $n=2$. The dashed and solid lines are from analytic and numeric calculations for $Y^{-1}$ respectively. They are in good agreement with each other. We see that the required $g_d$ for $\bar{f}f$ is much smaller than
$A'A'$ and $A'\phi$. This is because the $s$-channel resonant enhancement $Y^{-1}_{\rm res}$ is proportional to $g_d^2$. By contrast, for $A'A'$ and $A'\phi$, their $Y^{-1}$ are both proportional to $g_d^4$ and do not depend on $\epsilon$. 
The required $g_d $ for $A'\phi$ is smaller than $A'A'$ due to the accidental resonant enhancement factor $m_\psi / m_\phi \gg 1$.

In Fig.~\ref{fig:mDM-gd}, we also compare our transient results with zero temperature examples labeled as $T=0$ and drawn with thin dotted lines. For $A'A'$ and $A'\phi$ channels, we force $A'$ and $\phi$
to be massless. Unsurprisingly, the required $g_d$ is smaller for $T=0$ compared with transient annihilation for $A'A'$ channel. However, for $A'\phi$, the required $g_d$ increased for $T=0$ since there is no accidental resonant enhancement.
For $\bar{f}f$ channel, it does not need long resonant period 
comparing with the normal resonant annihilation, which fine-tuned to $m_{A',0} = 2 m_\psi$ due to slightly larger $g_d$. 
This relation can be understood from the relic abundance in Eq.~(\ref{eq:Ym1fin}), which is proportional to $g_d^{-2}\kappa^{1/n}$.
Therefore, a larger $g_d$  can compensate the shorter resonant period, which originated from the lower bound of $\kappa$.

\begin{figure}
	\centering
		\includegraphics[width=0.9\columnwidth]{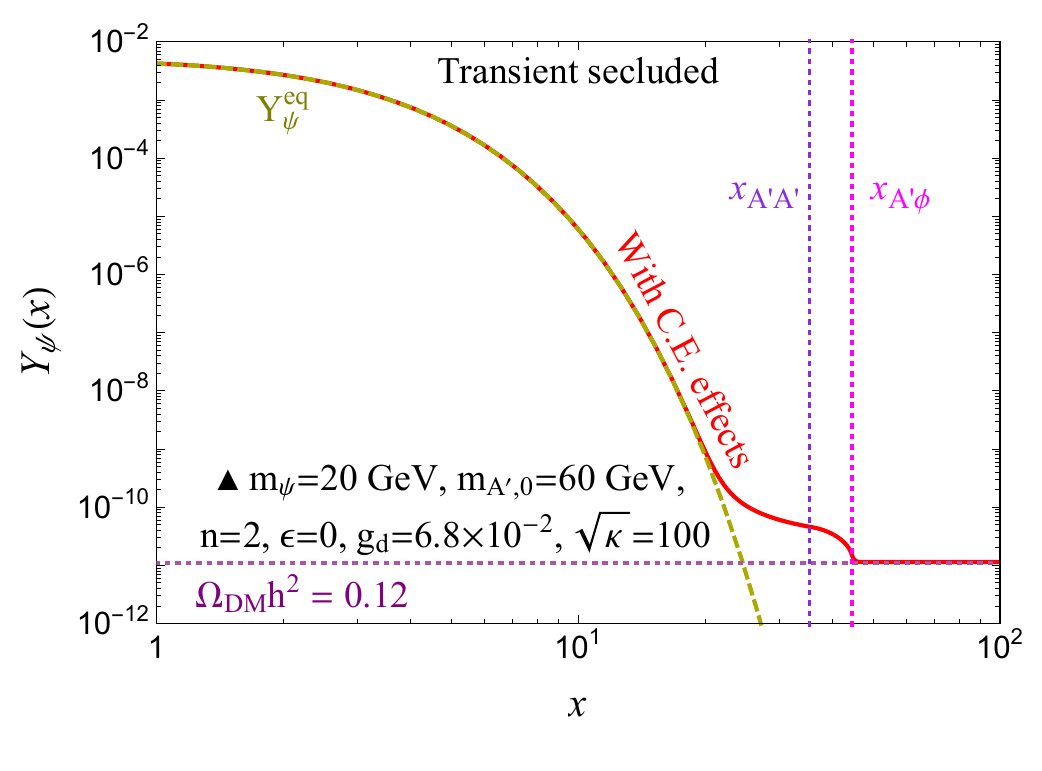} 
		\includegraphics[width=0.9\columnwidth]{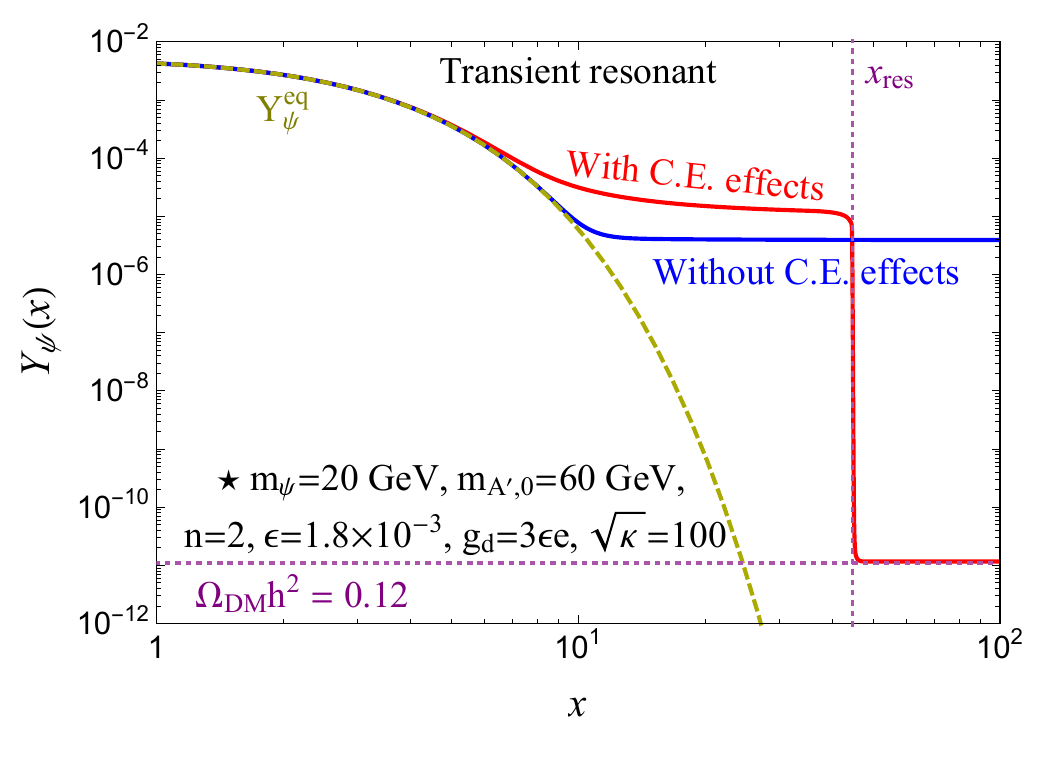}
	\caption{The DM yield $Y$ as a function of $x$ for all the transient annihilation channels
		$A'A'$, $A'\phi$ and $\bar{f}f$ included. Two benchmarks with large ($\blacktriangle$) and small ($\bigstar$) $g_d$ are shown respectively in the upper and lower panels for $n=2$. For the lower panel, the $Y$'s with and without cosmological evolution (C.E.) effects are both shown for comparison.}
	\label{fig:Y-x-plot}
\end{figure}

In Fig.~\ref{fig:Y-x-plot}, we plot $Y$ as a function of $x$ and show the evolution with all annihilation channels included. For the upper panel, it shows that the contribution from transient secluded annihilation $A'A'$ and $A' \phi$ dominates with large $g_d$ and can lead to the correct relic abundance. 
For the lower panel, the transient resonant annihilation $\bar{f}f$ dominates with small $g_d$. We show the two benchmarks with (without) cosmological evolution effects using $\sqrt{\kappa} = 100$ (0). Without evolution effects, the transient secluded channels are kinematically forbidden.
\\

\section{Constraints}

For transient secluded annihilations $A'A'$ and $A'\phi$, the collider and DD bounds can be easily evaded by choosing a tiny $\epsilon$. However, it does not work for the transient resonant annihilation $\bar{f}f$. This is because its annihilation cross-section will be proportional to $\epsilon^2$ in the small $\epsilon$ limit, making the cross-section not large enough to provide the right relic abundance. Therefore, we choose a moderate $\epsilon$ satisfying $g_d \lesssim \sqrt{n_{\rm dof}} \epsilon e$. Hence, the annihilation cross-section is still proportional to $g_d^2$. 
We consider the constraints from DD bounds \cite{XENON:2018voc, PandaX:2021osp} and dilepton and mono-photon searches at colliders~\cite{BESIII:2017fwv, BaBar:2017tiz, Zhang:2019wnz, LHCb:2019vmc, CMS:2019buh, ATLAS:2019erb}. Since the transient annihilations are either forbidden or weakened in the late universe, the indirect detection does not constrain the scenario in general. For example, the transient resonant benchmark has annihilation cross-section of about $10^{-34}{\rm cm}^3/{\rm s}$ at the CMB era which is much smaller than the CMB and indirect search bounds.

The nucleon scattering cross-section is given as,
\begin{align}
	\sigma_{p}^{\rm SI} = \frac{\epsilon^2 e^2 g_d^2}{\pi} \frac{\mu^2_{\psi p}}{m_{A',0}^4},
\end{align}
with reduced mass $\mu_{\psi p}\equiv m_\psi m_p/(m_\psi + m_p)$. Since the resonant cross-section is proportional to $g_d^2 \epsilon^2 e^2/( g_d^2 + n_{\rm dof} \epsilon^2 e^2)$ and the nucleon scattering cross-section is proportional to $g_d^2 \epsilon^2 e^2$, there is an optimal point around $g_d \approx \sqrt{n_{\rm dof}} \epsilon e$ yielding a small DD signal for a given DM relic abundance. We choose this point as our benchmark $\bigstar$ in Fig.~\ref{fig:Y-x-plot}.

\begin{figure}
	\centering
	\includegraphics[width=0.98 \columnwidth]{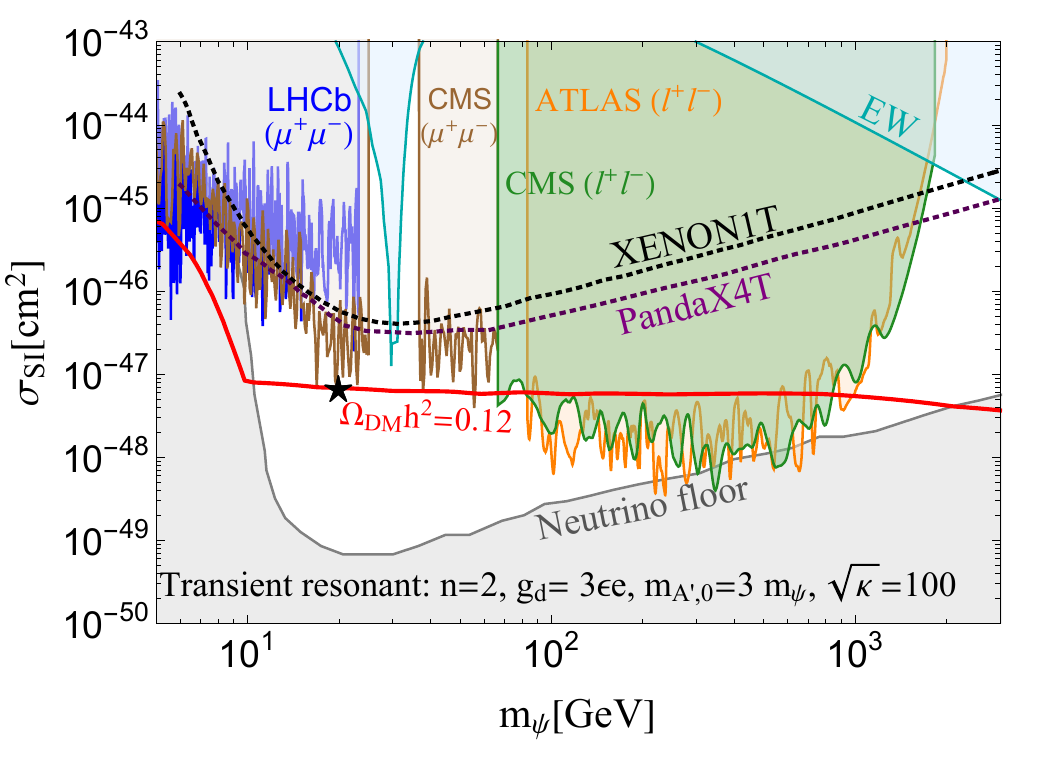} 
	\caption{The constraints for transient $s$-channel resonance model from DD~\cite{XENON:2018voc, PandaX:2021osp} and dilepton searches at LHC \cite{LHCb:2019vmc, CMS:2019buh, ATLAS:2019erb, CMS:2021ctt}. We choose optimal $g_d = 3 \epsilon e$ to evade the existing limits and fix the mass ratio $m_{A',0} = 3 m_\psi$ and $\sqrt{\kappa} = 100 $. The benchmark $\bigstar$ of Fig.~\ref{fig:Y-x-plot} is also displayed.}
	\label{fig:mDM-sigmaSI}
\end{figure}   
 
In Fig.~\ref{fig:mDM-sigmaSI}, we show the DD and collider constraints for transient resonant annihilation and compare them with the relic abundance requirement. The only free parameters are $\epsilon$ and $m_\psi$. The others are fixed by the relations $m_{A', 0}  = 3 m_\psi$, $g_d = 3 \epsilon e$ and $\sqrt{\kappa} = 100$ for $n=2$. The signal shown in red becomes flat for $m_\psi > 10 $ GeV because the relic abundance requires $g_d^2 / m_\psi^2$ to be roughly constant while $\sigma_{\rm SI} $
is proportional to $g_d^4/m_\psi^4$ in this setup.  

We can see that the transient resonant annihilation helps to evade the current DD limits from XENON1T and PandaX-4T even for $m_\psi\neq m_{A',0}/2$. Moreover, it is well within the reach of future experiments and can be soon tested before reaching the neutrino floor. 
The dilepton searches from LHCb \cite{LHCb:2019vmc}, CMS \cite{CMS:2019buh, CMS:2021ctt} and ATLAS \cite{ATLAS:2019erb} are most relevant and stringent; even with significant invisible decay branching ratio  $A' \to \bar{\psi}\psi$. The ATLAS and CMS results leave two windows for DM mass within $10 $--$100$ GeV and $\gtrsim 1$ TeV.
Besides the benchmark, taking a smaller $g_d / (\epsilon e)$ ratio will not help. With the required $g_d$ from the relic abundance, $\epsilon$ is not small enough to evade DD and dilepton searches.
On the other hand, choosing a large $g_d / (\epsilon e)$ ratio does not work either. The required $g_d$ for the relic abundance increases significantly; thus, $\epsilon $ is still too large. Taking a smaller mass ratio $r_0 \to 2$ will definitely help as shown in the $T=0$ example in Fig.~\ref{fig:mDM-gd}. This significantly decreases $g_d$; thus, alleviates the tension from $\epsilon$. 
Therefore, the parameter space for the transient resonant annihilation is pretty restricted and
future collider and DD searches can soon fully test it.
\\

\section{Conclusion}
We studied the effects of cosmological evolution on DM annihilation in the early Universe; especially for the gauge boson mediator. They can open new transient secluded and resonant annihilation channels which change the experimental constraints on the model parameter space. 
We have obtained the analytic forms of the relic abundance for each transient annihilation and they are in good agreement with full numeric calculations.
We choose $r_0 > 2$ as an example for all three channels.
The transient secluded annihilations become fully secluded; with no DD and collider signal and even negligible indirect detection signal.
The transient resonant annihilation is a viable and promising WIMP variant. It can evade the current DD and dilepton searches and can be fully tested by experiments in the near future.  
There are two windows opening for DM mass within $10$--$100$ GeV and around 1 TeV. If $r_0\to 2$, there are more parameter space opens for transient resonant annihilation.
For $1<r_0 <2$, the transient annihilations are still viable and important. The dark Higgs has to be much lighter than the gauge boson, which is a feature of the transient annihilation scenario.
\\

\begin{acknowledgments}
The authors would like to thank Chengcheng Han, Joachim Kopp, Lu Yin for helpful discussions, and Wen Han Chiu for proofreading. The work of JL is supported by National Science Foundation of China under Grant No. 12075005 and by Peking University under startup Grant No. 7101502458. The work of XPW is supported by National Science Foundation of China under Grant No. 12005009. KPX is supported by the University of Nebraska-Lincoln.
\end{acknowledgments}

\appendix
\section{Appendix}

\subsection{The general phase transition}\label{app:pt}

We briefly comment on the case of a general second-order phase transition (or crossover) whose temperature dependence of $m_{A'}$ is not in the simple $T^n$ form. In that case, our treatment in the main text still provides a generic methodology to deal with the transient scenarios. Taking the transient resonant channel as an example, the local properties of $A'$ smoothly crossing the critical point, i.e. $m_{A'}(T_{\rm res}) = 2m_\psi$, is sufficient for the transient resonant calculation; it is not necessary to know the global picture of function $m_{A'}(T)$.

The reason behind this argument is that the main transient resonant effect only lasts very short time ($x \in [44.6, 45.6]$), as shown in Fig.~\ref{fig:Y-x-plot} in main text and Fig.~\ref{fig:sigmav}. Therefore, it is always viable to make a Taylor expansion at the resonant temperature $T_{\rm res}$, simplifying the $m_{A'}(T)$ to a polynomial function of $T$,  
\begin{align}\label{S2}
	m_{A'}^2(T) = m_{A'}^2(T_{\rm res}) + \sum_{n=1} c_n \left(T -T_{\rm res}\right)^n ,
\end{align}
where the resonant temperature $T_{\rm res}$ and polynomial coefficients $c_n$ need to be determined.
Since the resonant time period is short, a finite $n$ is enough to describe $A'$ mass, e.g. $n =2$. Eq.~(\ref{S2}) can be rewritten to
\begin{align}
	m_{A'}^2(T) = m_{A',0}^2 + \sum_{n=1} f_n T^n ,
\end{align}
which can make use of the knowledge of the known zero temperature mass for $A'$. The coefficients $f_n$ should be given by the phase transition around the resonance, but the global information of the phase transition is not necessary for the transient resonance calculation. Then the subsequent calculation on relic abundance can be performed in a way very similar to that in our paper. 

In Eq.~(\ref{eq:Apmass}) and Eq.~(\ref{eq:Yres-short}), we have calculated the transient effect for each single $T^n$ term with arbitrary $n$.
As shown in the discussion below Eq.~(\ref{eq:Yres-short}), different $n$ does not change the qualitative feature of the mechanism, but just affect the choice of $\kappa$. Therefore, the methodology presented in our paper works for the generic case.

\subsection{Thermal corrections to the potential}\label{app:thermal}

We show a more detailed treatment for the thermal potential of $\phi$. The one-loop level thermal potential consists of the zero-temperature Coleman-Weinberg (CW) potential and the thermal integrals. The CW potential is given by the logarithmic terms, $\sum_i \frac{n_iM^4_i(\varphi)}{64\pi^2}\left(\ln\frac{M^2_i(\varphi)}{Q^2}-c_i\right)$, where $Q$ is renormalization scale, $n_i$ is degree of freedom of the $i$-th field, and $c_i$ = 3/2 (for scalar bosons and fermions), 5/6 (for gauge bosons). The thermal integrals are dominated by the light degrees of freedom, thus we can use the high temperature approximation. Combining the leading terms in the expansion and the CW potential together, we get
\begin{align}
	V(\phi,T)& =\frac{\mu_{d,0}^2+c_\phi T^2}{2}\phi^2+\frac{\lambda_d}{4}\phi^4  \\
	& + \sum_{\rm bosons}\frac{n_iM^4_i(\phi)}{64\pi^2}\left(\ln\frac{\alpha_BT^2}{Q^2}-c_i\right) \nonumber \\
	& +\sum_{\rm fermions}\frac{n_iM^4_i(\phi)}{64\pi^2}\left(\ln\frac{\alpha_FT^2}{Q^2}-c_i\right)+\cdots, \nonumber
\end{align}
where 
\begin{align}
	\log \alpha_B & =2\log4\pi -2\gamma_E+3/2, \nonumber \\
	\log\alpha_F & =2\log\pi -2\gamma_E+3/2, \nonumber 
\end{align}
and $\gamma_E$ is the Euler constant. We can see the CW part contributes to the scalar potential as a quartic coupling with weak (logarithmic) dependence on the temperature. Since the transient annihilations happen in a very short time period, as shown in Fig.~2 of the main text, the logarithmic part can be approximated as a constant and absorbed into the definition of $\lambda_d$. Therefore, the thermal potential can be written in the form of Eq.~(3). \\

\bibliographystyle{apsrev}
\bibliography{ref}

\begin{thebibliography}{60}
\expandafter\ifx\csname natexlab\endcsname\relax\def\natexlab#1{#1}\fi
\expandafter\ifx\csname bibnamefont\endcsname\relax
  \def\bibnamefont#1{#1}\fi
\expandafter\ifx\csname bibfnamefont\endcsname\relax
  \def\bibfnamefont#1{#1}\fi
\expandafter\ifx\csname citenamefont\endcsname\relax
  \def\citenamefont#1{#1}\fi
\expandafter\ifx\csname url\endcsname\relax
  \def\url#1{\texttt{#1}}\fi
\expandafter\ifx\csname urlprefix\endcsname\relax\def\urlprefix{URL }\fi
\providecommand{\bibinfo}[2]{#2}
\providecommand{\eprint}[2][]{\url{#2}}

\bibitem[{\citenamefont{Akerib et~al.}(2017)}]{LUX:2016ggv}
\bibinfo{author}{\bibfnamefont{D.~S.} \bibnamefont{Akerib}}
  \bibnamefont{et~al.} (\bibinfo{collaboration}{LUX}), \bibinfo{journal}{Phys.
  Rev. Lett.} \textbf{\bibinfo{volume}{118}}, \bibinfo{pages}{021303}
  (\bibinfo{year}{2017}), \eprint{1608.07648}.

\bibitem[{\citenamefont{Liu et~al.}(2019)}]{CDEX:2019hzn}
\bibinfo{author}{\bibfnamefont{Z.~Z.} \bibnamefont{Liu}} \bibnamefont{et~al.}
  (\bibinfo{collaboration}{CDEX}), \bibinfo{journal}{Phys. Rev. Lett.}
  \textbf{\bibinfo{volume}{123}}, \bibinfo{pages}{161301}
  (\bibinfo{year}{2019}), \eprint{1905.00354}.

\bibitem[{\citenamefont{Aprile et~al.}(2018)}]{XENON:2018voc}
\bibinfo{author}{\bibfnamefont{E.}~\bibnamefont{Aprile}} \bibnamefont{et~al.}
  (\bibinfo{collaboration}{XENON}), \bibinfo{journal}{Phys. Rev. Lett.}
  \textbf{\bibinfo{volume}{121}}, \bibinfo{pages}{111302}
  (\bibinfo{year}{2018}), \eprint{1805.12562}.

\bibitem[{\citenamefont{Meng et~al.}(2021)}]{PandaX:2021osp}
\bibinfo{author}{\bibfnamefont{Y.}~\bibnamefont{Meng}} \bibnamefont{et~al.}
  (\bibinfo{collaboration}{PandaX}) (\bibinfo{year}{2021}),
  \eprint{2107.13438}.

\bibitem[{\citenamefont{Aguilar et~al.}(2014)}]{AMS:2014xys}
\bibinfo{author}{\bibfnamefont{M.}~\bibnamefont{Aguilar}} \bibnamefont{et~al.}
  (\bibinfo{collaboration}{AMS}), \bibinfo{journal}{Phys. Rev. Lett.}
  \textbf{\bibinfo{volume}{113}}, \bibinfo{pages}{121102}
  (\bibinfo{year}{2014}).

\bibitem[{\citenamefont{Accardo et~al.}(2014)}]{AMS:2014bun}
\bibinfo{author}{\bibfnamefont{L.}~\bibnamefont{Accardo}} \bibnamefont{et~al.}
  (\bibinfo{collaboration}{AMS}), \bibinfo{journal}{Phys. Rev. Lett.}
  \textbf{\bibinfo{volume}{113}}, \bibinfo{pages}{121101}
  (\bibinfo{year}{2014}).

\bibitem[{\citenamefont{Ackermann et~al.}(2015)}]{Fermi-LAT:2015att}
\bibinfo{author}{\bibfnamefont{M.}~\bibnamefont{Ackermann}}
  \bibnamefont{et~al.} (\bibinfo{collaboration}{Fermi-LAT}),
  \bibinfo{journal}{Phys. Rev. Lett.} \textbf{\bibinfo{volume}{115}},
  \bibinfo{pages}{231301} (\bibinfo{year}{2015}), \eprint{1503.02641}.

\bibitem[{\citenamefont{Albert et~al.}(2017)}]{Fermi-LAT:2016uux}
\bibinfo{author}{\bibfnamefont{A.}~\bibnamefont{Albert}} \bibnamefont{et~al.}
  (\bibinfo{collaboration}{Fermi-LAT, DES}), \bibinfo{journal}{Astrophys. J.}
  \textbf{\bibinfo{volume}{834}}, \bibinfo{pages}{110} (\bibinfo{year}{2017}),
  \eprint{1611.03184}.

\bibitem[{\citenamefont{Ambrosi et~al.}(2017)}]{DAMPE:2017fbg}
\bibinfo{author}{\bibfnamefont{G.}~\bibnamefont{Ambrosi}} \bibnamefont{et~al.}
  (\bibinfo{collaboration}{DAMPE}), \bibinfo{journal}{Nature}
  \textbf{\bibinfo{volume}{552}}, \bibinfo{pages}{63} (\bibinfo{year}{2017}),
  \eprint{1711.10981}.

\bibitem[{\citenamefont{Sirunyan et~al.}(2019)}]{CMS:2018yfx}
\bibinfo{author}{\bibfnamefont{A.~M.} \bibnamefont{Sirunyan}}
  \bibnamefont{et~al.} (\bibinfo{collaboration}{CMS}), \bibinfo{journal}{Phys.
  Lett. B} \textbf{\bibinfo{volume}{793}}, \bibinfo{pages}{520}
  (\bibinfo{year}{2019}), \eprint{1809.05937}.

\bibitem[{\citenamefont{Aaboud et~al.}(2019)}]{ATLAS:2019cid}
\bibinfo{author}{\bibfnamefont{M.}~\bibnamefont{Aaboud}} \bibnamefont{et~al.}
  (\bibinfo{collaboration}{ATLAS}), \bibinfo{journal}{Phys. Rev. Lett.}
  \textbf{\bibinfo{volume}{122}}, \bibinfo{pages}{231801}
  (\bibinfo{year}{2019}), \eprint{1904.05105}.

\bibitem[{\citenamefont{Aad et~al.}(2021)}]{ATLAS:2021kxv}
\bibinfo{author}{\bibfnamefont{G.}~\bibnamefont{Aad}} \bibnamefont{et~al.}
  (\bibinfo{collaboration}{ATLAS}), \bibinfo{journal}{Phys. Rev. D}
  \textbf{\bibinfo{volume}{103}}, \bibinfo{pages}{112006}
  (\bibinfo{year}{2021}), \eprint{2102.10874}.

\bibitem[{\citenamefont{Tumasyan et~al.}(2021)}]{CMS:2021far}
\bibinfo{author}{\bibfnamefont{A.}~\bibnamefont{Tumasyan}} \bibnamefont{et~al.}
  (\bibinfo{collaboration}{CMS}) (\bibinfo{year}{2021}), \eprint{2107.13021}.

\bibitem[{\citenamefont{Holdom}(1986)}]{Holdom:1985ag}
\bibinfo{author}{\bibfnamefont{B.}~\bibnamefont{Holdom}},
  \bibinfo{journal}{Phys. Lett. B} \textbf{\bibinfo{volume}{166}},
  \bibinfo{pages}{196} (\bibinfo{year}{1986}).

\bibitem[{\citenamefont{Pospelov et~al.}(2008)\citenamefont{Pospelov, Ritz, and
  Voloshin}}]{Pospelov:2007mp}
\bibinfo{author}{\bibfnamefont{M.}~\bibnamefont{Pospelov}},
  \bibinfo{author}{\bibfnamefont{A.}~\bibnamefont{Ritz}}, \bibnamefont{and}
  \bibinfo{author}{\bibfnamefont{M.~B.} \bibnamefont{Voloshin}},
  \bibinfo{journal}{Phys. Lett. B} \textbf{\bibinfo{volume}{662}},
  \bibinfo{pages}{53} (\bibinfo{year}{2008}), \eprint{0711.4866}.

\bibitem[{\citenamefont{D'Agnolo and Ruderman}(2015)}]{DAgnolo:2015ujb}
\bibinfo{author}{\bibfnamefont{R.~T.} \bibnamefont{D'Agnolo}} \bibnamefont{and}
  \bibinfo{author}{\bibfnamefont{J.~T.} \bibnamefont{Ruderman}},
  \bibinfo{journal}{Phys. Rev. Lett.} \textbf{\bibinfo{volume}{115}},
  \bibinfo{pages}{061301} (\bibinfo{year}{2015}), \eprint{1505.07107}.

\bibitem[{\citenamefont{Cline et~al.}(2017)\citenamefont{Cline, Liu, Slatyer,
  and Xue}}]{Cline:2017tka}
\bibinfo{author}{\bibfnamefont{J.~M.} \bibnamefont{Cline}},
  \bibinfo{author}{\bibfnamefont{H.}~\bibnamefont{Liu}},
  \bibinfo{author}{\bibfnamefont{T.}~\bibnamefont{Slatyer}}, \bibnamefont{and}
  \bibinfo{author}{\bibfnamefont{W.}~\bibnamefont{Xue}},
  \bibinfo{journal}{Phys. Rev. D} \textbf{\bibinfo{volume}{96}},
  \bibinfo{pages}{083521} (\bibinfo{year}{2017}), \eprint{1702.07716}.

\bibitem[{\citenamefont{Fitzpatrick et~al.}(2020)\citenamefont{Fitzpatrick,
  Liu, Slatyer, and Tsai}}]{Fitzpatrick:2020vba}
\bibinfo{author}{\bibfnamefont{P.~J.} \bibnamefont{Fitzpatrick}},
  \bibinfo{author}{\bibfnamefont{H.}~\bibnamefont{Liu}},
  \bibinfo{author}{\bibfnamefont{T.~R.} \bibnamefont{Slatyer}},
  \bibnamefont{and} \bibinfo{author}{\bibfnamefont{Y.-D.} \bibnamefont{Tsai}}
  (\bibinfo{year}{2020}), \eprint{2011.01240}.

\bibitem[{\citenamefont{Xing and Zhu}(2021)}]{Xing:2021pkb}
\bibinfo{author}{\bibfnamefont{C.-Y.} \bibnamefont{Xing}} \bibnamefont{and}
  \bibinfo{author}{\bibfnamefont{S.-H.} \bibnamefont{Zhu}},
  \bibinfo{journal}{Phys. Rev. Lett.} \textbf{\bibinfo{volume}{127}},
  \bibinfo{pages}{061101} (\bibinfo{year}{2021}), \eprint{2102.02447}.

\bibitem[{\citenamefont{Fitzpatrick et~al.}(2021)\citenamefont{Fitzpatrick,
  Liu, Slatyer, and Tsai}}]{Fitzpatrick:2021cij}
\bibinfo{author}{\bibfnamefont{P.~J.} \bibnamefont{Fitzpatrick}},
  \bibinfo{author}{\bibfnamefont{H.}~\bibnamefont{Liu}},
  \bibinfo{author}{\bibfnamefont{T.~R.} \bibnamefont{Slatyer}},
  \bibnamefont{and} \bibinfo{author}{\bibfnamefont{Y.-D.} \bibnamefont{Tsai}}
  (\bibinfo{year}{2021}), \eprint{2105.05255}.

\bibitem[{\citenamefont{Cirelli et~al.}(2017)\citenamefont{Cirelli, Panci,
  Petraki, Sala, and Taoso}}]{Cirelli:2016rnw}
\bibinfo{author}{\bibfnamefont{M.}~\bibnamefont{Cirelli}},
  \bibinfo{author}{\bibfnamefont{P.}~\bibnamefont{Panci}},
  \bibinfo{author}{\bibfnamefont{K.}~\bibnamefont{Petraki}},
  \bibinfo{author}{\bibfnamefont{F.}~\bibnamefont{Sala}}, \bibnamefont{and}
  \bibinfo{author}{\bibfnamefont{M.}~\bibnamefont{Taoso}},
  \bibinfo{journal}{JCAP} \textbf{\bibinfo{volume}{05}}, \bibinfo{pages}{036}
  (\bibinfo{year}{2017}), \eprint{1612.07295}.

\bibitem[{\citenamefont{Liu et~al.}(2017)\citenamefont{Liu, Wang, and
  Yu}}]{Liu:2017lpo}
\bibinfo{author}{\bibfnamefont{J.}~\bibnamefont{Liu}},
  \bibinfo{author}{\bibfnamefont{X.-P.} \bibnamefont{Wang}}, \bibnamefont{and}
  \bibinfo{author}{\bibfnamefont{F.}~\bibnamefont{Yu}}, \bibinfo{journal}{JHEP}
  \textbf{\bibinfo{volume}{06}}, \bibinfo{pages}{077} (\bibinfo{year}{2017}),
  \eprint{1704.00730}.

\bibitem[{\citenamefont{Tucker-Smith and Weiner}(2001)}]{Tucker-Smith:2001myb}
\bibinfo{author}{\bibfnamefont{D.}~\bibnamefont{Tucker-Smith}}
  \bibnamefont{and} \bibinfo{author}{\bibfnamefont{N.}~\bibnamefont{Weiner}},
  \bibinfo{journal}{Phys. Rev. D} \textbf{\bibinfo{volume}{64}},
  \bibinfo{pages}{043502} (\bibinfo{year}{2001}), \eprint{hep-ph/0101138}.

\bibitem[{\citenamefont{Alekhin et~al.}(2016)}]{Alekhin:2015byh}
\bibinfo{author}{\bibfnamefont{S.}~\bibnamefont{Alekhin}} \bibnamefont{et~al.},
  \bibinfo{journal}{Rept. Prog. Phys.} \textbf{\bibinfo{volume}{79}},
  \bibinfo{pages}{124201} (\bibinfo{year}{2016}), \eprint{1504.04855}.

\bibitem[{\citenamefont{Battaglieri
  et~al.}(2017{\natexlab{a}})}]{Battaglieri:2017aum}
\bibinfo{author}{\bibfnamefont{M.}~\bibnamefont{Battaglieri}}
  \bibnamefont{et~al.}, in \emph{\bibinfo{booktitle}{{U.S. Cosmic Visions: New
  Ideas in Dark Matter}}} (\bibinfo{year}{2017}{\natexlab{a}}),
  \eprint{1707.04591}.

\bibitem[{\citenamefont{Battaglieri et~al.}(2017{\natexlab{b}})}]{BDX:2017jub}
\bibinfo{author}{\bibfnamefont{M.}~\bibnamefont{Battaglieri}}
  \bibnamefont{et~al.} (\bibinfo{collaboration}{BDX})
  (\bibinfo{year}{2017}{\natexlab{b}}), \eprint{1712.01518}.

\bibitem[{\citenamefont{Berlin et~al.}(2019)\citenamefont{Berlin, Blinov,
  Krnjaic, Schuster, and Toro}}]{Berlin:2018bsc}
\bibinfo{author}{\bibfnamefont{A.}~\bibnamefont{Berlin}},
  \bibinfo{author}{\bibfnamefont{N.}~\bibnamefont{Blinov}},
  \bibinfo{author}{\bibfnamefont{G.}~\bibnamefont{Krnjaic}},
  \bibinfo{author}{\bibfnamefont{P.}~\bibnamefont{Schuster}}, \bibnamefont{and}
  \bibinfo{author}{\bibfnamefont{N.}~\bibnamefont{Toro}},
  \bibinfo{journal}{Phys. Rev. D} \textbf{\bibinfo{volume}{99}},
  \bibinfo{pages}{075001} (\bibinfo{year}{2019}), \eprint{1807.01730}.

\bibitem[{\citenamefont{\r{A}kesson et~al.}(2018)}]{LDMX:2018cma}
\bibinfo{author}{\bibfnamefont{T.}~\bibnamefont{\r{A}kesson}}
  \bibnamefont{et~al.} (\bibinfo{collaboration}{LDMX}) (\bibinfo{year}{2018}),
  \eprint{1808.05219}.

\bibitem[{\citenamefont{Tsai et~al.}(2021)\citenamefont{Tsai, deNiverville, and
  Liu}}]{Tsai:2019buq}
\bibinfo{author}{\bibfnamefont{Y.-D.} \bibnamefont{Tsai}},
  \bibinfo{author}{\bibfnamefont{P.}~\bibnamefont{deNiverville}},
  \bibnamefont{and} \bibinfo{author}{\bibfnamefont{M.~X.} \bibnamefont{Liu}},
  \bibinfo{journal}{Phys. Rev. Lett.} \textbf{\bibinfo{volume}{126}},
  \bibinfo{pages}{181801} (\bibinfo{year}{2021}), \eprint{1908.07525}.

\bibitem[{\citenamefont{Slatyer}(2016)}]{Slatyer:2015kla}
\bibinfo{author}{\bibfnamefont{T.~R.} \bibnamefont{Slatyer}},
  \bibinfo{journal}{Phys. Rev. D} \textbf{\bibinfo{volume}{93}},
  \bibinfo{pages}{023521} (\bibinfo{year}{2016}), \eprint{1506.03812}.

\bibitem[{\citenamefont{Cohen et~al.}(2008)\citenamefont{Cohen, Morrissey, and
  Pierce}}]{Cohen:2008nb}
\bibinfo{author}{\bibfnamefont{T.}~\bibnamefont{Cohen}},
  \bibinfo{author}{\bibfnamefont{D.~E.} \bibnamefont{Morrissey}},
  \bibnamefont{and} \bibinfo{author}{\bibfnamefont{A.}~\bibnamefont{Pierce}},
  \bibinfo{journal}{Phys. Rev. D} \textbf{\bibinfo{volume}{78}},
  \bibinfo{pages}{111701} (\bibinfo{year}{2008}), \eprint{0808.3994}.

\bibitem[{\citenamefont{Baker and Kopp}(2017)}]{Baker:2016xzo}
\bibinfo{author}{\bibfnamefont{M.~J.} \bibnamefont{Baker}} \bibnamefont{and}
  \bibinfo{author}{\bibfnamefont{J.}~\bibnamefont{Kopp}},
  \bibinfo{journal}{Phys. Rev. Lett.} \textbf{\bibinfo{volume}{119}},
  \bibinfo{pages}{061801} (\bibinfo{year}{2017}), \eprint{1608.07578}.

\bibitem[{\citenamefont{Kobakhidze et~al.}(2018)\citenamefont{Kobakhidze,
  Schmidt, and Talia}}]{Kobakhidze:2017ini}
\bibinfo{author}{\bibfnamefont{A.}~\bibnamefont{Kobakhidze}},
  \bibinfo{author}{\bibfnamefont{M.~A.} \bibnamefont{Schmidt}},
  \bibnamefont{and} \bibinfo{author}{\bibfnamefont{M.}~\bibnamefont{Talia}},
  \bibinfo{journal}{Phys. Rev. D} \textbf{\bibinfo{volume}{98}},
  \bibinfo{pages}{095026} (\bibinfo{year}{2018}), \eprint{1712.05170}.

\bibitem[{\citenamefont{Baker et~al.}(2018)\citenamefont{Baker, Breitbach,
  Kopp, and Mittnacht}}]{Baker:2017zwx}
\bibinfo{author}{\bibfnamefont{M.~J.} \bibnamefont{Baker}},
  \bibinfo{author}{\bibfnamefont{M.}~\bibnamefont{Breitbach}},
  \bibinfo{author}{\bibfnamefont{J.}~\bibnamefont{Kopp}}, \bibnamefont{and}
  \bibinfo{author}{\bibfnamefont{L.}~\bibnamefont{Mittnacht}},
  \bibinfo{journal}{JHEP} \textbf{\bibinfo{volume}{03}}, \bibinfo{pages}{114}
  (\bibinfo{year}{2018}), \eprint{1712.03962}.

\bibitem[{\citenamefont{Baker and Mittnacht}(2019)}]{Baker:2018vos}
\bibinfo{author}{\bibfnamefont{M.~J.} \bibnamefont{Baker}} \bibnamefont{and}
  \bibinfo{author}{\bibfnamefont{L.}~\bibnamefont{Mittnacht}},
  \bibinfo{journal}{JHEP} \textbf{\bibinfo{volume}{05}}, \bibinfo{pages}{070}
  (\bibinfo{year}{2019}), \eprint{1811.03101}.

\bibitem[{\citenamefont{Hektor et~al.}(2018)\citenamefont{Hektor, Kannike, and
  Vaskonen}}]{Hektor:2018esx}
\bibinfo{author}{\bibfnamefont{A.}~\bibnamefont{Hektor}},
  \bibinfo{author}{\bibfnamefont{K.}~\bibnamefont{Kannike}}, \bibnamefont{and}
  \bibinfo{author}{\bibfnamefont{V.}~\bibnamefont{Vaskonen}},
  \bibinfo{journal}{Phys. Rev. D} \textbf{\bibinfo{volume}{98}},
  \bibinfo{pages}{015032} (\bibinfo{year}{2018}), \eprint{1801.06184}.

\bibitem[{\citenamefont{Bian and Tang}(2018)}]{Bian:2018mkl}
\bibinfo{author}{\bibfnamefont{L.}~\bibnamefont{Bian}} \bibnamefont{and}
  \bibinfo{author}{\bibfnamefont{Y.-L.} \bibnamefont{Tang}},
  \bibinfo{journal}{JHEP} \textbf{\bibinfo{volume}{12}}, \bibinfo{pages}{006}
  (\bibinfo{year}{2018}), \eprint{1810.03172}.

\bibitem[{\citenamefont{Bian and Liu}(2019)}]{Bian:2018bxr}
\bibinfo{author}{\bibfnamefont{L.}~\bibnamefont{Bian}} \bibnamefont{and}
  \bibinfo{author}{\bibfnamefont{X.}~\bibnamefont{Liu}},
  \bibinfo{journal}{Phys. Rev. D} \textbf{\bibinfo{volume}{99}},
  \bibinfo{pages}{055003} (\bibinfo{year}{2019}), \eprint{1811.03279}.

\bibitem[{\citenamefont{Kobakhidze et~al.}(2020)\citenamefont{Kobakhidze,
  Schmidt, and Talia}}]{Kobakhidze:2019tts}
\bibinfo{author}{\bibfnamefont{A.}~\bibnamefont{Kobakhidze}},
  \bibinfo{author}{\bibfnamefont{M.~A.} \bibnamefont{Schmidt}},
  \bibnamefont{and} \bibinfo{author}{\bibfnamefont{M.}~\bibnamefont{Talia}},
  \bibinfo{journal}{JCAP} \textbf{\bibinfo{volume}{03}}, \bibinfo{pages}{059}
  (\bibinfo{year}{2020}), \eprint{1910.01433}.

\bibitem[{\citenamefont{Heurtier and Partouche}(2020)}]{Heurtier:2019beu}
\bibinfo{author}{\bibfnamefont{L.}~\bibnamefont{Heurtier}} \bibnamefont{and}
  \bibinfo{author}{\bibfnamefont{H.}~\bibnamefont{Partouche}},
  \bibinfo{journal}{Phys. Rev. D} \textbf{\bibinfo{volume}{101}},
  \bibinfo{pages}{043527} (\bibinfo{year}{2020}), \eprint{1912.02828}.

\bibitem[{\citenamefont{Darm\'e et~al.}(2019)\citenamefont{Darm\'e, Hryczuk,
  Karamitros, and Roszkowski}}]{Darme:2019wpd}
\bibinfo{author}{\bibfnamefont{L.}~\bibnamefont{Darm\'e}},
  \bibinfo{author}{\bibfnamefont{A.}~\bibnamefont{Hryczuk}},
  \bibinfo{author}{\bibfnamefont{D.}~\bibnamefont{Karamitros}},
  \bibnamefont{and}
  \bibinfo{author}{\bibfnamefont{L.}~\bibnamefont{Roszkowski}},
  \bibinfo{journal}{JHEP} \textbf{\bibinfo{volume}{11}}, \bibinfo{pages}{159}
  (\bibinfo{year}{2019}), \eprint{1908.05685}.

\bibitem[{\citenamefont{Davoudiasl and Mohlabeng}(2020)}]{Davoudiasl:2019xeb}
\bibinfo{author}{\bibfnamefont{H.}~\bibnamefont{Davoudiasl}} \bibnamefont{and}
  \bibinfo{author}{\bibfnamefont{G.}~\bibnamefont{Mohlabeng}},
  \bibinfo{journal}{JHEP} \textbf{\bibinfo{volume}{04}}, \bibinfo{pages}{177}
  (\bibinfo{year}{2020}), \eprint{1912.05572}.

\bibitem[{\citenamefont{De~Romeri et~al.}(2020)\citenamefont{De~Romeri,
  Karamitros, Lebedev, and Toma}}]{DeRomeri:2020wng}
\bibinfo{author}{\bibfnamefont{V.}~\bibnamefont{De~Romeri}},
  \bibinfo{author}{\bibfnamefont{D.}~\bibnamefont{Karamitros}},
  \bibinfo{author}{\bibfnamefont{O.}~\bibnamefont{Lebedev}}, \bibnamefont{and}
  \bibinfo{author}{\bibfnamefont{T.}~\bibnamefont{Toma}},
  \bibinfo{journal}{JHEP} \textbf{\bibinfo{volume}{10}}, \bibinfo{pages}{137}
  (\bibinfo{year}{2020}), \eprint{2003.12606}.

\bibitem[{\citenamefont{Jaramillo et~al.}(2021)\citenamefont{Jaramillo,
  Lindner, and Rodejohann}}]{Jaramillo:2020dde}
\bibinfo{author}{\bibfnamefont{C.}~\bibnamefont{Jaramillo}},
  \bibinfo{author}{\bibfnamefont{M.}~\bibnamefont{Lindner}}, \bibnamefont{and}
  \bibinfo{author}{\bibfnamefont{W.}~\bibnamefont{Rodejohann}},
  \bibinfo{journal}{JCAP} \textbf{\bibinfo{volume}{04}}, \bibinfo{pages}{023}
  (\bibinfo{year}{2021}), \eprint{2004.12904}.

\bibitem[{\citenamefont{Croon et~al.}(2020)\citenamefont{Croon, Elor, Houtz,
  Murayama, and White}}]{Croon:2020ntf}
\bibinfo{author}{\bibfnamefont{D.}~\bibnamefont{Croon}},
  \bibinfo{author}{\bibfnamefont{G.}~\bibnamefont{Elor}},
  \bibinfo{author}{\bibfnamefont{R.}~\bibnamefont{Houtz}},
  \bibinfo{author}{\bibfnamefont{H.}~\bibnamefont{Murayama}}, \bibnamefont{and}
  \bibinfo{author}{\bibfnamefont{G.}~\bibnamefont{White}}
  (\bibinfo{year}{2020}), \eprint{2012.15284}.

\bibitem[{\citenamefont{Nakayama and Yin}(2021)}]{Nakayama:2021avl}
\bibinfo{author}{\bibfnamefont{K.}~\bibnamefont{Nakayama}} \bibnamefont{and}
  \bibinfo{author}{\bibfnamefont{W.}~\bibnamefont{Yin}} (\bibinfo{year}{2021}),
  \eprint{2105.14549}.

\bibitem[{\citenamefont{Batell and Ghalsasi}(2021)}]{Batell:2021ofv}
\bibinfo{author}{\bibfnamefont{B.}~\bibnamefont{Batell}} \bibnamefont{and}
  \bibinfo{author}{\bibfnamefont{A.}~\bibnamefont{Ghalsasi}}
  (\bibinfo{year}{2021}), \eprint{2109.04476}.

\bibitem[{\citenamefont{Davoudiasl et~al.}(2004)\citenamefont{Davoudiasl,
  Kitano, Kribs, Murayama, and Steinhardt}}]{Davoudiasl:2004gf}
\bibinfo{author}{\bibfnamefont{H.}~\bibnamefont{Davoudiasl}},
  \bibinfo{author}{\bibfnamefont{R.}~\bibnamefont{Kitano}},
  \bibinfo{author}{\bibfnamefont{G.~D.} \bibnamefont{Kribs}},
  \bibinfo{author}{\bibfnamefont{H.}~\bibnamefont{Murayama}}, \bibnamefont{and}
  \bibinfo{author}{\bibfnamefont{P.~J.} \bibnamefont{Steinhardt}},
  \bibinfo{journal}{Phys. Rev. Lett.} \textbf{\bibinfo{volume}{93}},
  \bibinfo{pages}{201301} (\bibinfo{year}{2004}), \eprint{hep-ph/0403019}.

\bibitem[{\citenamefont{Caldwell and Gubser}(2013)}]{Caldwell:2013mox}
\bibinfo{author}{\bibfnamefont{R.~R.} \bibnamefont{Caldwell}} \bibnamefont{and}
  \bibinfo{author}{\bibfnamefont{S.~S.} \bibnamefont{Gubser}},
  \bibinfo{journal}{Phys. Rev. D} \textbf{\bibinfo{volume}{87}},
  \bibinfo{pages}{063523} (\bibinfo{year}{2013}), \eprint{1302.1201}.

\bibitem[{\citenamefont{Dine et~al.}(1995)\citenamefont{Dine, Randall, and
  Thomas}}]{Dine:1995uk}
\bibinfo{author}{\bibfnamefont{M.}~\bibnamefont{Dine}},
  \bibinfo{author}{\bibfnamefont{L.}~\bibnamefont{Randall}}, \bibnamefont{and}
  \bibinfo{author}{\bibfnamefont{S.~D.} \bibnamefont{Thomas}},
  \bibinfo{journal}{Phys. Rev. Lett.} \textbf{\bibinfo{volume}{75}},
  \bibinfo{pages}{398} (\bibinfo{year}{1995}), \eprint{hep-ph/9503303}.

\bibitem[{\citenamefont{Dine et~al.}(1996)\citenamefont{Dine, Randall, and
  Thomas}}]{Dine:1995kz}
\bibinfo{author}{\bibfnamefont{M.}~\bibnamefont{Dine}},
  \bibinfo{author}{\bibfnamefont{L.}~\bibnamefont{Randall}}, \bibnamefont{and}
  \bibinfo{author}{\bibfnamefont{S.~D.} \bibnamefont{Thomas}},
  \bibinfo{journal}{Nucl. Phys. B} \textbf{\bibinfo{volume}{458}},
  \bibinfo{pages}{291} (\bibinfo{year}{1996}), \eprint{hep-ph/9507453}.

\bibitem[{\citenamefont{Tang}(2016)}]{Tang:2016mot}
\bibinfo{author}{\bibfnamefont{Y.}~\bibnamefont{Tang}}, \bibinfo{journal}{Phys.
  Lett. B} \textbf{\bibinfo{volume}{757}}, \bibinfo{pages}{387}
  (\bibinfo{year}{2016}), \eprint{1603.00165}.

\bibitem[{\citenamefont{Binder et~al.}(2021)\citenamefont{Binder, Bringmann,
  Gustafsson, and Hryczuk}}]{Binder:2021bmg}
\bibinfo{author}{\bibfnamefont{T.}~\bibnamefont{Binder}},
  \bibinfo{author}{\bibfnamefont{T.}~\bibnamefont{Bringmann}},
  \bibinfo{author}{\bibfnamefont{M.}~\bibnamefont{Gustafsson}},
  \bibnamefont{and} \bibinfo{author}{\bibfnamefont{A.}~\bibnamefont{Hryczuk}},
  \bibinfo{journal}{Eur. Phys. J. C} \textbf{\bibinfo{volume}{81}},
  \bibinfo{pages}{577} (\bibinfo{year}{2021}), \eprint{2103.01944}.

\bibitem[{\citenamefont{Ablikim et~al.}(2017)}]{BESIII:2017fwv}
\bibinfo{author}{\bibfnamefont{M.}~\bibnamefont{Ablikim}} \bibnamefont{et~al.}
  (\bibinfo{collaboration}{BESIII}), \bibinfo{journal}{Phys. Lett. B}
  \textbf{\bibinfo{volume}{774}}, \bibinfo{pages}{252} (\bibinfo{year}{2017}),
  \eprint{1705.04265}.

\bibitem[{\citenamefont{Lees et~al.}(2017)}]{BaBar:2017tiz}
\bibinfo{author}{\bibfnamefont{J.~P.} \bibnamefont{Lees}} \bibnamefont{et~al.}
  (\bibinfo{collaboration}{BaBar}), \bibinfo{journal}{Phys. Rev. Lett.}
  \textbf{\bibinfo{volume}{119}}, \bibinfo{pages}{131804}
  (\bibinfo{year}{2017}), \eprint{1702.03327}.

\bibitem[{\citenamefont{Zhang et~al.}(2019)\citenamefont{Zhang, Zhang, Song,
  Pan, Niu, and Li}}]{Zhang:2019wnz}
\bibinfo{author}{\bibfnamefont{Y.}~\bibnamefont{Zhang}},
  \bibinfo{author}{\bibfnamefont{W.-T.} \bibnamefont{Zhang}},
  \bibinfo{author}{\bibfnamefont{M.}~\bibnamefont{Song}},
  \bibinfo{author}{\bibfnamefont{X.-A.} \bibnamefont{Pan}},
  \bibinfo{author}{\bibfnamefont{Z.-M.} \bibnamefont{Niu}}, \bibnamefont{and}
  \bibinfo{author}{\bibfnamefont{G.}~\bibnamefont{Li}}, \bibinfo{journal}{Phys.
  Rev. D} \textbf{\bibinfo{volume}{100}}, \bibinfo{pages}{115016}
  (\bibinfo{year}{2019}), \eprint{1907.07046}.

\bibitem[{\citenamefont{Aaij et~al.}(2020)}]{LHCb:2019vmc}
\bibinfo{author}{\bibfnamefont{R.}~\bibnamefont{Aaij}} \bibnamefont{et~al.}
  (\bibinfo{collaboration}{LHCb}), \bibinfo{journal}{Phys. Rev. Lett.}
  \textbf{\bibinfo{volume}{124}}, \bibinfo{pages}{041801}
  (\bibinfo{year}{2020}), \eprint{1910.06926}.

\bibitem[{\citenamefont{Sirunyan et~al.}(2020)}]{CMS:2019buh}
\bibinfo{author}{\bibfnamefont{A.~M.} \bibnamefont{Sirunyan}}
  \bibnamefont{et~al.} (\bibinfo{collaboration}{CMS}), \bibinfo{journal}{Phys.
  Rev. Lett.} \textbf{\bibinfo{volume}{124}}, \bibinfo{pages}{131802}
  (\bibinfo{year}{2020}), \eprint{1912.04776}.

\bibitem[{\citenamefont{Aad et~al.}(2019)}]{ATLAS:2019erb}
\bibinfo{author}{\bibfnamefont{G.}~\bibnamefont{Aad}} \bibnamefont{et~al.}
  (\bibinfo{collaboration}{ATLAS}), \bibinfo{journal}{Phys. Lett. B}
  \textbf{\bibinfo{volume}{796}}, \bibinfo{pages}{68} (\bibinfo{year}{2019}),
  \eprint{1903.06248}.

\bibitem[{\citenamefont{Sirunyan et~al.}(2021)}]{CMS:2021ctt}
\bibinfo{author}{\bibfnamefont{A.~M.} \bibnamefont{Sirunyan}}
  \bibnamefont{et~al.} (\bibinfo{collaboration}{CMS}), \bibinfo{journal}{JHEP}
  \textbf{\bibinfo{volume}{07}}, \bibinfo{pages}{208} (\bibinfo{year}{2021}),
  \eprint{2103.02708}.

\end{thebibliography}

\end{document}